\documentstyle[12pt]{article}

\newcommand{\be}{\begin{equation}}
\newcommand{\ee}{\end{equation}}
\newcommand{\bea}{\begin{eqnarray}}
\newcommand{\eea}{\end{eqnarray}}

\newcommand{\alg}{{\cal A}}

\newcommand{\eqn}[1]{(\ref{#1})}
\newcommand{\complex}{{\bb C}} %% complex numbers
\newcommand{\zed}{{\bb Z}} %% integers
\newcommand{\real}{{\bb R}} %% real numbers
\newcommand{\complexs}{{\bbs C}} %% small complex numbers
\newcommand{\id}{{\bb I}} %% identity operator
\newcommand{\torus}{{\bb T}} %% nctorus
\newcommand{\NO}{\,\mbox{$\circ\atop\circ$}\,} % Normal ordering
\def\nn{\nonumber}

\def\slash{\!\!\!/}
\def\slashs{\!\!\!\!/}
\def\Dirac{{D\!\!\!\!/\,}}

\font\mybb=msbm10 at 12pt
\def\bb#1{\hbox{\mybb#1}}
\font\mybbs=msbm10 at 9pt
\def\bbs#1{\hbox{\mybbs#1}}

\def\e{{\rm e}}

\makeatletter
\newdimen\normalarrayskip              % skip between lines
\newdimen\minarrayskip                 % minimal skip between lines
\normalarrayskip\baselineskip
\minarrayskip\jot
\newif\ifold             \oldtrue            \def\new{\oldfalse}
\makeatother

\newlength{\extraspace}
\newlength{\extraspaces}
\begin{document}

\addtolength{\baselineskip}{.8mm}

\renewcommand{\footnotesize}{\small}

\thispagestyle{empty}

\begin{flushright}
\baselineskip=12pt
NBI-HE-98-43\\
hep-th/9812235\\
\hfill{  }\\ December 1998
\end{flushright}
\vspace{.5cm}

\begin{center}

\baselineskip=24pt

{\Large\bf{Spectral Geometry of Heterotic Compactifications}}\\[15mm]

\baselineskip=12pt

{\bf David D. Song} \\[3mm]
{\it Department of Physics\\University of Oxford\\Clarendon Laboratory, Parks
Road, Oxford OX1 3PU, U.K.\\{\tt d.song1@physics.oxford.ac.uk}}
\\[6mm]
{\bf Richard J.\ Szabo}\\[3mm]
{\it The Niels Bohr Institute\\ Blegdamsvej 17, DK-2100\\ Copenhagen \O,
Denmark\\{\tt szabo@nbi.dk}}\\[40mm]

{\sc Abstract}

\begin{center}

\begin{minipage}{14cm}

\baselineskip=12pt

The structure of heterotic string target space compactifications is studied
using the formalism of the noncommutative geometry associated with lattice
vertex operator algebras. The spectral triples of the noncommutative spacetimes
are constructed and used to show that the intrinsic gauge field degrees of
freedom disappear in the low-energy sectors of these spacetimes. The quantum
geometry is thereby determined in much the same way as for ordinary superstring
target spaces. In this setting, non-abelian gauge theories on the classical
spacetimes arise from the $K$-theory of the effective target spaces.

\end{minipage}
\end{center}

\end{center}

\vfill
\newpage

\pagestyle{plain}
\setcounter{page}{1}

The description of spacetime at very short distance scales where the effects of
quantum gravity become important is a complex problem. The conventional ideas
of geometry and general relativity are expected to break down at scales of the
order of the Planck length of the spacetime, so that one can anticipate very
different physics in this regime. Target space duality in string theory (see
\cite{giveon2} for a review) implies that very small distances are unobservable
and leads to the notion of quantum geometry which is the modification of
ordinary, classical geometry suitable for the physics implied by string theory.
A natural tool to explore these ideas is noncommutative geometry \cite{ncg}. In
this approach, spacetime is described by the spectral triple ${\cal T} = ({\cal
A},{\cal H},D)$, where ${\cal A}$ is a $*$-algebra of bounded operators acting
on a separable Hilbert space ${\cal H}$, and $D$ is an appropriately defined
generalized Dirac operator on ${\cal H}$ which defines a metric on the
spacetime. A Riemannian spin manifold $M$ is reconstructed by choosing the
abelian $*$-algebra ${\cal A} = C^\infty(M,\complex)$ of smooth complex-valued
functions on $M$ which acts on the Hilbert space ${\cal H} = L^2(M,S)$ of
square-integrable spinors on $M$ by pointwise multiplication. The Riemannian
geometry of the manifold is then described by the usual Dirac operator
$D=i\,\nabla\slashs$ whose square is the inverse of the invariant line element
of $M$.

Compactifications of string theory have been described in
\cite{fg}--\cite{lizzi3} using this formalism based on the noncommutative space
constructed from the $*$-algebra of worldsheet vertex operators. These spaces
naturally encode the quantum geometry implied by string duality. An effective
spectral action for superstring oscillatory modes has been constructed in
\cite{chams} and a manifestly duality-symmetric action for the low-energy modes
in \cite{lizzi3}. In this paper we will apply this formalism to describe some
aspects of the effective target space geometry associated with a heterotic
string theory \cite{gross2}--\cite{giveon1}. Heterotic strings are
characterized by the fact that they are chiral and their left-moving sector
describes the usual superstrings, while their right-moving sector is
non-supersymmetric and contains a coupling of extra internal bosonic fields to
gauge degrees of freedom. This gives a natural way of introducing non-abelian
gauge symmetry into the noncommutative string spacetime by a sort of
Kaluza-Klein mechanism. After constructing the spectral triple $\cal T$
naturally associated to a heterotic string compactification,\footnote{One
should also introduce a $\zed_2$ grading and a real structure which define,
respectively, a Hochschild cycle and a Poincar\'e dual for the geometry. The
even real spectral triples associated with lattice vertex operator algebras are
described in detail in \cite{lizzi1,lizzi3}. In the following we shall only
need the first three basic elements of $\cal T$.} we shall examine the large
distance limit of the geometry and show that the internal gauge degrees of
freedom are unobservable with the same effective classical spacetime as in the
case of the ordinary superstring arising. In particular, the quantum geometry
relates the same inequivalent classical spacetimes. This is in some contrast to
the expectations from string theory \cite{giveon1} in which the low-energy
effective theory is altered by the presence of internal gauge degrees of
freedom. We will show that the latter effects automatically turn the
compactification into a finite-dimensional noncommutative space, so that the
internal modes of the strings describe very small wrapped directions of the
effective spacetime. We will also see that these internal degrees of freedom
are observable in the classical spacetime when one considers gauge theories as
induced by noncommutative $K$-theory. The latter structure illuminates the
manner in which the gauge field modes curl up into very small extra dimensions
to form a noncommutative space at very short distances but are nevertheless
observable in the physics of the macroscopic spacetime.

We consider closed string propagation on a $d$-dimensional torus
$T^d=\real^d/2\pi\Lambda$, where $\Lambda$ is a Euclidean lattice of rank $d$
with metric $G_{mn}$ and antisymmetric tensor $B_{mn}$. We are interested in
vacuum spacetime configurations, and so we shall always take the background
fields to be constant. We fix a basis $e_m$ of $\Lambda$. The winding numbers
$w^me_m$ of the strings wrapping around the cycles of the torus live in
$\Lambda$ while their momenta $p_me^m$ lie in the dual lattice $\Lambda^*$,
which has basis $e^m$ dual to $e_m$ and metric $G^{mn}$ with
$G^{mn}G_{nl}=\delta_l^m$, and label string windings around the cycles of the
dual torus $T_*^d=\real^d/2\pi\Lambda^*$. The internal gauge degrees of freedom
of the string are described by a simply-laced finite-dimensional Lie group
$\cal G$. The root lattice $\Gamma_{\cal G}$ of $\cal G$ is an even self-dual
Euclidean lattice of rank $8k$ whose non-zero roots all have length
$\sqrt2$.\footnote{$\cal G$ is therefore some combination of $SO(2n)$, $SU(n)$,
$E_6$, $E_7$ or $E_8$. In the standard construction of the heterotic string,
$k=2$ and there are two unique root lattices $\Gamma_8\oplus\Gamma_8$ and
$\Gamma_{16}$ of the groups $E_8\times E_8$ and $Spin(32)/\zed_2$,
respectively.} It has constant metric $G_{MN}$ and antisymmetric tensor
$B_{MN}$.\footnote{In this paper lower case Latin indices $m=1,\dots,d$ label
components in $\Lambda$, upper case Latin indices $M=1,\dots,8k$ the components
in $\Gamma_{\cal G}$, and Greek indices $\mu=1,\dots,d+8k$ label the directions
of $\Lambda\oplus\Gamma_{\cal G}$ and $\Lambda^*\oplus\Gamma_{\cal G}$. An
implicit sum over repeated upper and lower indices is always assumed, with
indices raised and lowered by the appropriate metric.} We fix a basis $e_M$ of
$\Gamma_{\cal G}$. The constant gauge fields on the strings are given by maps
$A:\Gamma_{\cal G}\to\Lambda^*$ whose lattice frame components are
$A=A_m^N\,e^m\otimes e_N$. The strings wrap around the cycles of the isospin
space $T^{8k}=\real^{8k}/2\pi\Gamma_{\cal G}$ with winding number
$w^Me_M\in\Gamma_{\cal G}$ and momentum $p_Me^M\in\Gamma_{\cal G}$.

The Narain lattice \cite{narain1} is defined as
\be
\Lambda_{\rm N}=\Lambda^*\oplus\Lambda\oplus\Gamma_{\cal G}\oplus\Gamma_{\cal
G}
\label{lattice}\ee
On \eqn{lattice} we define a $\zed$-bilinear form $\langle\cdot,\cdot\rangle$
with respect to the standard basis by $\langle e_m,e_n\rangle=2G_{mn}$,
$\langle e_M,e_N\rangle=2G_{MN}$, and $\langle e_m,e_N\rangle=0$. With this
definition $\Lambda_{\rm N}$ is an even self-dual Lorentzian lattice of rank
$2d+16k$ which can be thought of as defining a $(d+8k,d+8k)$ dimensional
Lorentzian torus $T^d\times T_*^d\times T^{8k}\times
T^{8k}=\real^{d+8k}\times\real^{d+8k}/2\pi\Lambda_{\rm N}$. We will use the
basis
\be
u_\mu =\left(e_m+\mbox{$\frac{1}{2}$}A_m^Le_L\right)\oplus e_M
\label{basis}\ee
of $\Lambda\oplus\Gamma_{\cal G}$ whose dual basis is
\be
u^{\mu} =e^{m}\oplus\left(-\mbox{$\frac{1}{2}$}A_m^M e^{m}+e^{M}\right)
\label{dualbasis}\ee
With respect to the basis $u_\mu\oplus u^\nu$, the metric tensor of
\eqn{lattice} is $2g_{\mu\nu}\oplus2g^{\lambda\rho}$ and the antisymmetric
tensor is $2\beta_{\mu\nu}\oplus2\beta^{\lambda\rho}$, where
\be
g_{\mu\nu}=\left( \begin{array}{cc}
          \left(G+\frac{1}{4}A^L\otimes A_L\right)_{mn} & \frac{1}{2}A_{mN} \\
           \frac{1}{2} A_{Mn} & G_{MN}
           \end{array} \right)
{}~~~~~~,~~~~~~\beta_{\mu\nu}=\left( \begin{array}{cc}
               B_{mn} & \frac{1}{2} A_{mN} \\
               -\frac{1}{2}A_{Mn} & B_{MN}
                \end{array} \right)
\label{g}\ee
The background matrices are then defined as $d_{\mu\nu}^{\pm} = g _{\mu\nu}\pm
\beta_{\mu\nu}$, so that $d_{MN}^+$ coincides with the Cartan matrix of $\cal
G$.

Given the left and right chiral momenta $p_{\mu}^{\pm} =\frac{1}{\sqrt
2}(p_{\mu} \pm d_{\mu\nu}^{\pm} w^{\nu})$ with respect to the standard basis of
$\Lambda_{\rm N}$, we define a mapping ${\cal P}:\Lambda_{\rm N}\to\Lambda_{\rm
N}$ by
\be
{\cal P}\{p_{\mu}^{\pm}\}={\cal P}\{p_l^{\pm}\}\oplus{\cal
P}\{p_L^{\pm}\}=u^{\nu}\,p_{\nu}^{\pm}
\label{calPdef}\ee
The Narain lattice vectors \eqn{calPdef} can be written in terms of the
standard basis of $\Lambda_{\rm N}$ as
\bea
{\cal P}\{p_l^+\} &=& \left[p_m + w^n\left(G+B-\mbox{$\frac{1}{4}$}\,A^L\otimes
A_L\right)_{nm} -\mbox{$\frac{1}{2}$}\left(p_N + w^M
(G+B)_{MN}\right)A_m^N\right]e^m\nn\\& &\label{pl+}\\{\cal P}\{p_L^+\} &=&
\left[p_N+w^M(G+B)_{MN} + A_{Nn} w^n\right] e^N \label{pL+}\\{\cal P}\{p_l^-\}
&=& \left[p_m - w^n\left(G-B-\mbox{$\frac{1}{4}$}\,A^L\otimes A_L\right)_{nm}
-\mbox{$\frac{1}{2}$}\left(p_N + w^M (G+B)_{MN}\right)A_m^N \right]e^m\nn\\&
&\label{pl-}\\{\cal P}\{p_L^-\}&=& \left[p_N-w^M(G-B)_{MN} \right]e^N
\label{pL-}\eea
We see that with the phase space constraints
\be
p_N=d_{MN}^-\,w^M
\label{phaseconstr}\ee
the right-moving modes (\ref{pL-}) of the isospin space vanish. With this
truncation of modes, the vectors (\ref{pl+})--(\ref{pl-}) generate the
$(d+8k,d)$ dimensional even self-dual Lorentzian lattice
\be
\Pi^-\Lambda_{\rm N}=\Lambda^*\oplus\Lambda\oplus\Gamma_{\cal G}
\label{latticeproj}\ee
where $\Pi^-$ is the orthogonal projection onto the subspace defined by
\eqn{phaseconstr}. The lattice \eqn{latticeproj} defines the heterotic string
compactification which is isomorphic to $T^d\times T_*^d\times T^{8k}$.

The spectral geometry of the heterotic string spacetime can now be constructed
using the truncation method above. Namely, we may use the construction of
\cite{lizzi1} to construct a spectral triple for the ordinary bosonic string
spacetime determined by the Narain lattice \eqn{lattice}, and then we project
onto the chiral heterotic subspace using the projection operator $\Pi^-$. The
classical embedding functions corresponding to the heterotic lattice
\eqn{latticeproj} are in this way given by the mode expansions
\bea
X_+^{\nu} (z_+) &=& x_+^{\nu} +i\left\langle u^{\nu}\,,\,{\cal
P}\{p_{\mu}^+\}\right\rangle\log z_+ + \sum_{k\not= 0}
\frac{1}{ik}\,\alpha_k^{(+)\nu}\,z_+^{-k} \label{XI+}\\
X_-^{n} (z_-) &=& x_-^{n}+i\left\langle u^n\,,\,{\cal
P}\{p_m^-\}\right\rangle\log z_- + \sum_{k\not= 0}
\frac{1}{ik}\,\alpha_k^{(-)n}\,z_-^{-k}
\label{XI-}\eea
where $x_\pm^m\in T^d\times T_*^d$, $x_+^M\in T^{8k}$,
$z_\pm=\e^{-i(\tau\pm\sigma)}$ and $(\tau,\sigma)\in\real\times S^1$. Upon
quantization the position and momentum zero modes $x_\pm$ and $p^\pm$ form a
canonically conjugate pair in each chiral sector, while the oscillator modes
generate the standard Heisenberg algebras with respect to the metric
$g^{\mu\nu}$ in the left moving sector and with respect to $g^{mn}$ in the
right moving sector. Fixing spin structures on the tori, the zero modes act on
spaces of square integrable spinors while the non-zero modes act on a mutually
commuting pair of Fock spaces ${\cal F}^\pm$. The operators \eqn{XI+} and
\eqn{XI-} therefore act on the Hilbert space
\be
{\cal H}=L^2(T^d\times T_*^d,S)\otimes_\complexs L^2(T^{8k},S_{\cal
G})\otimes{\cal F}^+\{\alpha_k^{(+)\mu}\}\otimes{\cal F}^-\{\alpha_k^{(-)m}\}
\label{HX}\ee
where $S\to T^d\times T_*^d$ and $S_{\cal G}\to T^{8k}$ are the respective spin
bundles. The spinor module ${\cal S}=C^\infty(T^d\times T_*^d,S)$ is a dense
subspace of $L^2(T^d\times T_*^d,S)$ which has a natural $\zed_2$-grading
${\cal S}={\cal S}^+\oplus{\cal S}^-$ coming from the chirality grading of the
corresponding Clifford bundle that acts irreducibly on $S$. This grading splits
the Hilbert space \eqn{HX} orthogonally into chiral and antichiral subspaces as
${\cal H}={\cal H}^+\oplus{\cal H}^-$. The module ${\cal S}$ thereby carries an
irreducible representation of the double toroidal Clifford algebra
$\{\gamma_m^\pm,\gamma_n^\pm\}=\pm2g_{mn}$ (with all other anticommutators
vanishing). We shall take all gamma-matrices to be Hermitian. Likewise, ${\cal
S}_{\cal G}=C^\infty(T^{8k},S_{\cal G})$ is a dense subspace of
$L^2(T^{8k},S_{\cal G})$ which carries an irreducible representation of the
Clifford algebra $\{\gamma_M,\gamma_N\}=2G_{MN}$ of the internal space
$T^{8k}$.

We may now construct an appropriate $*$-algebra for the noncommutative
heterotic spacetime. The necessary algebra is the vertex operator algebra
$\alg$ which acts densely on \eqn{HX} and diagonally with respect to its
chirality decomposition, and is constructed using the operator-state
correspondence. Given $q_m^{\pm} = \frac{1}{\sqrt 2}(q_m\pm
d_{m\mu}^{\pm}v^{\mu})\in\Lambda^*\oplus\Lambda$ and $K\in\Gamma_{\cal G}$, the
$*$-algebra $\alg$ is generated by the basic tachyon vertex
operators\footnote{Note that the operators \eqn{Vnl} and \eqn{VN} are unbounded
on $\cal H$ and should be more precisely defined, for instance by the usual
quantum field theoretical smearing \cite{susyncg} or alternatively by the
cutoff definition introduced in \cite{lizzi3}.}
\bea
V_{q^+q^-}(z_+,z_-)&=&c_{q^+q^-}\{p_n^+,p_l^-\}\,\NO\e^{-iq_m^+
X_+^m(z_+)-iq_{m}^-
X_-^{m}(z_-)}\NO\label{Vnl}\\E_K(z_+)&=&c_K\{p_N^+\}\,
\NO\e^{-iK_MX_+^M(z_+)}\NO
\label{VN}\eea
where $\NO\cdot\NO$ denotes the standard Wick normal ordering of operators. The
vertex operators \eqn{Vnl} are the usual ones for toroidal compactifications
with the operator-valued phases $c_{q^+q^-}\{p_n^+,p_l^-\}=\e^{i\pi q_mw^m}$.
In the purely bosonic case, they generate the inner automorphism group of the
vertex operator algebra which is generically an enhancement of the affine
$U(1)_+^d\times U(1)_-^d$ gauge symmetry of the string spacetime \cite{lizzi5}.
In the heterotic case, the interesting new feature which arises are the vertex
operators \eqn{VN} associated with the isospin space. The cocycles
$c_K\{p_N^+\}$ in \eqn{VN} are defined by expanding the internal lattice vector
$K_M=k_N(e^N)_M$ in a basis of $\Gamma_{\cal G}$, with $k_N\in\zed$, and
defining an ordered product $\star:\Gamma_{\cal G}\times\Gamma_{\cal G}\to\zed$
by
\be
K\star K'=\sum_{N>M}G^{LP}\,k_Nk'_M\,(e^N)_L(e^M)_P
\label{stardef}\ee
Then the operator-valued one-cocycles are defined as
\be
c_K\{p_N^+\}=\e^{i\pi\,p^+\star K}
\label{cocycledef}\ee
and they generate the algebra
\be
c_K\{p_N^+-K_N'\}\,c_{K'}\{p_N^+\}=\varepsilon(K,K')\,c_{K+K'}\{p_N^+\}
\label{cocycleprod}\ee
where
\be
\varepsilon(K,K')=\e^{i\pi\,K\star K'}
\label{2cocycle}\ee
is a $\zed_2$-valued two-cocycle of the root lattice of $\cal G$. It follows
that the operators
\be
H^M(z_+)=-iz_+\,\partial_{z_+}X_+^M(z_+)=\sum_{k=-\infty}^\infty
\alpha_k^{(+)M}\,z_+^{-k}~~~~~~{\rm with}~~\alpha_0^{(+)M}=g^{M\mu}\,p_\mu^+
\label{HMdef}\ee
and $E_K(z_+)$ for $K\in\Gamma_{\cal G}$ generate a level 1 representation of
the current algebra associated with the Lie group $\cal G$ in the Chevalley
basis \cite{gross2,voa}. Thus the internal gauge symmetry of the strings is
also represented as an inner automorphism of the heterotic spacetime.

The final ingredient we require are the Dirac operators. The guiding principle
in constructing them \cite{fg}--\cite{susyncg},\cite{lizzi2,chamspec} is to
find an appropriate supersymmetric extension of the underlying field theory and
to take the Dirac operators to be the projections of the corresponding
supercharges onto the fermionic zero-modes in the Ramond sector of the model.
This relates the Dirac operators to the generators of target space
reparametrization symmetries. In the case at hand, the right-moving sector of
the underlying worldsheet sigma-model is supersymmetric, and so we only need to
supersymmetrize the left-moving sector. This is tantamount to considering more
symmetric background matrices
\be
\tilde d^+_{\mu\nu}\equiv\tilde g_{\mu\nu}+\tilde\beta_{\mu\nu}=\left(
\begin{array}{cc}
                  \left(G+\frac{1}{4}A^L\otimes A_L\right)_{mn}+B_{mn}  &
A_{mN} \\
                  A_{Mn} &  G_{MN}+B_{MN}
                 \end{array} \right)
\ee
of the Narain lattice \eqn{lattice} representing the addition of left-moving
fermion fields. The corresponding $N=1$ supercharges are then
\be
Q^{\pm}(z_\pm)=-\mbox{$\frac{i}{\sqrt 2}$}\,\tilde
g_{\mu\nu}\,\psi_{\pm}^{\mu}(z_\pm)\,\partial_{z_\pm}X_\pm^{\nu}(z_\pm)
\ee
where $\psi_{\pm}^{\mu}(z_\pm)$ are Majorana-Weyl spinor fields. Upon
quantization, the fermionic zero-modes of the Ramond sector of the worldsheet
theory may be identified with Dirac algebra generators as
$\psi_\pm^{(0)m}=\sqrt2\,\tilde g^{mn}\gamma_n^\pm$ and
$\psi_+^{(0)M}=\sqrt2\,\tilde g^{MN}\gamma_N$. If $\Pi_{\rm R}^{(0)}$ denotes
the corresponding projection operator acting on the Hilbert space of the
supersymmetric sigma-model, then the Dirac operators are defined by
\bea
\Dirac^+(z_+) &\equiv&\Pi^-\,\Pi_{\rm R}^{(0)}\,Q^+(z_+)\,\Pi_{\rm
R}^{(0)}\,\Pi^-~=~\gamma_M\otimes
H^M(z_+)+\sum_{k=-\infty}^{\infty}\gamma_m^+\otimes\alpha_k^{(+)m}
\,z_+^{-k}\label{D+}\\\Dirac^{-}(z_{-})&\equiv&\Pi^-\,\Pi_{\rm R}^{(0)}
\,Q^-(z_-)\,\Pi_{\rm R}^{(0)}\,\Pi^-~=~\sum_{k=-\infty}^{\infty}
\gamma_{m}^-\otimes\alpha_k^{(-)m}\,z_-^{-k}
\label{D-}\eea
where $\alpha_0^{(\pm)m}=g^{m\mu}\,p_\mu^\pm$. These Dirac operators are
associated with the conserved currents of the diffeomorphism symmetry of the
effective target space $T^d\times T_*^d\times T^{8k}$. We shall use the
chirally symmetric and antisymmetric combinations $\Dirac=\Dirac^++\Dirac^-$
and $\overline{\Dirac}=\Dirac^+-\Dirac^-$. The noncommutative geometry of the
heterotic string spacetime is then encoded in the spectral triples $(\alg,{\cal
H},\Dirac)$ and $(\alg,{\cal H},\overline{\Dirac})$.

The first feature of the heterotic spectral triples we shall explore is the
emergence of classical spacetime. This arises from the kernel of the given
Dirac operator \cite{lizzi1}, which defines the low-energy projection of the
string background such that the spacetime is given as a special case of the
spectral action principle of noncommutative geometry
\cite{chamspec,specaction}. After some algebra it can be shown that these
kernels each decompose into $2^d$ subspaces of the Hilbert space $\cal H$
according to
\bea
\ker\Dirac&\cong&\left[\bigotimes_{m=1}^d\left(\overline{\cal
H}_+^{(m)}\oplus\overline{\cal H}_-^{(m)}\right)\right]\otimes{\cal H}_{\rm
int}\label{kerD}\\\ker\overline{\Dirac}&\cong&\left[\bigotimes_{m=1}^d
\left({\cal H}_+^{(m)}\oplus{\cal H}_-^{(m)}\right)\right]
\otimes{\cal H}_{\rm int}
\label{kerbarD}\eea
The subspace ${\cal H}_{\rm int}=\bigotimes_M\ker H^M$ is defined by the zero
modes of the Cartan generators of the internal gauge symmetry. It consists of
the projection of all Fock states onto the vacuum, and from the constraint
\eqn{phaseconstr} it follows that its $L^2$ states have the quantum number
restrictions
\be
\left[\delta^M_N+\mbox{$\frac14$}\left(A^l\otimes
A_l\right)^M_N\right]w^N=-\mbox{$\frac14$}\,A^{Mn}\biggl[p_n+
\left(G+B+\mbox{$\frac14$}\,A^L\otimes A_L\right)_{nm}\,w^m\biggr]
\label{Hintdef}\ee
The chiral and antichiral subspaces $\overline{\cal H}_\pm^{(m)}$ are defined
for each $m=1,\dots,d$ by again projecting onto the Fock vacuum (thereby
suppressing the oscillatory modes of the strings) and by imposing conditions on
the spinor representations defined via the actions on them by the
gamma-matrices. After some algebra we find that these conditions are
equivalent, respectively, to $L^2$ quantum number constraints as
$g^{m\nu}d_{\nu\lambda}^+\gamma_m^+=g^{m\nu}d_{\nu\lambda}^-\gamma_m^-\iff
p_m=0$ and $\gamma_m^+=-\gamma_m^-\iff w^m=0$. Similarly, in ${\cal
H}_\pm^{(m)}$ we find, respectively, the analogous conditions
$\gamma_m^+=\gamma_m^-\iff w^m=0$ and
$g^{m\nu}d_{\nu\lambda}^+\gamma_m^+=-g^{m\nu}d_{\nu\lambda}^-\gamma_m^-\iff
p_m=0$. It is straightforward to see that these conditions imply that
\be
p_M=w^M=0~~~~~~\forall M=1,\dots,8k
\label{intvanish}\ee
and so the internal string windings corresponding to the gauge degrees of
freedom are suppressed. Defining
\be
\tilde\gamma_m^\pm\equiv G^{ln}\left(G\pm B\mp\mbox{$\frac14$}\,A^L\otimes
A_L\right)_{nm}\,\gamma_l^\pm
\label{tildegammadef}\ee
we then have $\tilde\gamma_m^+=\tilde\gamma_m^-$ in $\overline{\cal H}_+^{(m)}$
and $\tilde\gamma_m^+=-\tilde\gamma_m^-$ in ${\cal H}_-^{(m)}$.

The various subspaces in \eqn{kerD} and \eqn{kerbarD} are all isomorphic and
correspond to different choices of spin structure on the $d$-tori
\cite{lizzi1}, for example the isomorphism $\overline{\cal
H}_+^{(m)}\leftrightarrow\overline{\cal H}_-^{(n)}$ is given by
$\tilde\gamma_m^\pm\leftrightarrow\pm\gamma_n^\pm$. They define classical
spacetimes represented by low-energy spectral triples. Let us fix the
completely antichiral subspace
\be
\overline{\cal H}_0^{(-)}=\overline{\cal
H}_-^{(1)}\otimes\cdots\otimes\overline{\cal H}_-^{(d)}
\label{antichiraldef}\ee
of \eqn{kerD}. In this subspace the spinors are antiperiodic along the elements
of a homology basis so that the associated spin bundle is completely twisted,
$\gamma_m^+=-\gamma_m^-\equiv\sigma_m$, and $w^m=0$, for all $m=1,\dots,d$. Let
$\overline{\Pi}_0^{(-)}:{\cal H}\to\overline{\cal H}_0^{(-)}$ be the
corresponding orthogonal projection. From \eqn{Hintdef} and \eqn{intvanish} it
then follows that the internal gauge degrees of freedom of the string vanish,
\be
A^M_m=0
\label{Avanish}\ee
According to the operator-state correspondence, the subalgebra of vertex
operators acting on \eqn{antichiraldef} is given by the projection of the
commutant of $\Dirac$ in $\alg$,
\be
\overline{\alg}_0^{(-)}=\overline{\Pi}_0^{(-)}\Bigl({\rm
End}_\Dirac\,\alg\Bigr)\,\overline{\Pi}_0^{(-)}
\label{barA0def}\ee
which is the largest subalgebra of $\alg$ with the property
$\overline{\alg}_0^{(-)}\,\overline{\cal H}_0^{(-)}=\overline{\cal H}_0^{(-)}$.
Since
\bea
\overline{\Pi}_0^{(-)}\,\Bigl[\Dirac\,,\,\id\otimes
V_{q^+q^-}\Bigr]\,\overline{\Pi}_0^{(-)}&=&g^{mn}\,\sigma_m\otimes
\left(q_n^+-q_n^-\right)\,\overline{\Pi}_0^{(-)}\,V_{q^+q^-}\,
\overline{\Pi}_0^{(-)}\label{DVq}\\\overline{\Pi}_0^{(-)}\,
\Bigl[\Dirac\,,\,\id\otimes E_K\Bigr]\,\overline{\Pi}_0^{(-)}&=&g^{MN}\,
\gamma_M\otimes K_N\,\overline{\Pi}_0^{(-)}\,E_K\,\overline{\Pi}_0^{(-)}
\label{DEK}\eea
it follows that the generators of \eqn{barA0def} have zero oscillatory modes,
zero internal momentum $K_N=0~~\forall N$ and zero winding numbers
$v^m=0~~\forall m$. It also implies that all internal momentum and winding
modes vanish, $q_M=v^M=0~~\forall M$. We therefore have an isomorphism
$\overline{\alg}_0^{(-)}\cong C^\infty(T^d,\complex)$ of $*$-algebras.
Furthermore, it is straightforward to compute the restriction of the Dirac
operator $\overline{\Dirac}$ to \eqn{antichiraldef} to find
\be
i\,\partial\slash\equiv\overline{\Dirac}~\overline{\Pi}_0^{(-)}
=-i\,G^{mn}\,\sigma_m\otimes\mbox{$\frac\partial{\partial x^n}$}
\label{barDrestrict}\ee
where $x^m=\frac1{\sqrt2}(x_+^m+x_-^m)\in T^d$.

This shows that the low-energy projection of the spectral triple associated
with the Dirac operator $\overline{\Dirac}$ is given by
\be
\left(\overline{\alg}_0^{(-)}~,~\overline{\cal
H}_0^{(-)}~,~\overline{\Dirac}~\overline{\Pi}_0^{(-)}\right)~\cong~
\biggl(C^\infty(T^d,\complex)~,~L^2(T^d,S^-)~,~i\,\partial\slash\biggr)
\label{spectripproj}\ee
where $S^-\to T^d$ is the spin bundle defined by the projection onto the spinor
boundary conditions of \eqn{antichiraldef}. The spectral triple
\eqn{spectripproj} describes the ordinary manifold $T^d$ with its flat metric
$G_{mn}$. Naively, one expects \cite{giveon1} the low-energy limit of the model
to be affected by the gauge field $A:\Gamma_{\cal G}\to\Lambda^*$ which mixes
the target space and internal lattice degrees of freedom. This is because it
appears in the cross terms $d_{mM}^\pm$ of the background matrices and it also
effectively shifts the metric of $T^d$ as $G_{mn}\to
G_{mn}+\frac14A^L_nA_{Lm}$. However, the definition of classical spacetime in
the noncommutative geometry formalism has effectively taken the extreme limit
whereby the compactification radii are very large ($G_{mn}\gg A^L_nA_{Lm}$).
The low-energy regime emerges as the subspace whereby all internal gauge
degrees of freedom are suppressed and only the classical geometry of the torus
$T^d$ is observable. This is precisely what one would anticipate upon observing
the string spacetime at very large distance scales.

The quantum geometry of the manifold \eqn{spectripproj} comes about from the
observation \cite{lizzi1} that the change in Dirac operator
$\Dirac\leftrightarrow\overline{\Dirac}$ corresponds to a change of metric on
the manifold, so that, by general covariance, the two corresponding spectral
triples are isomorphic, $(\alg,{\cal H},\overline{\Dirac})\cong(\alg,{\cal
H},\Dirac)$. There are several unitary transformations $T:{\cal H}\to{\cal H}$
which implement this isomorphism by defining automorphisms of the vertex
operator algebra and mapping the two Dirac operators into one another as
$\overline{\Dirac}\,T=T\,\Dirac$. They can be obtained by examining the web of
isomorphisms between the subspace \eqn{antichiraldef} and the $2^d$ subspaces
in \eqn{kerbarD}. We shall only sketch here how the duality group appears --
the full construction can be carried out as described in \cite{lizzi1}. Let us
first consider the completely antichiral subspace of \eqn{kerbarD}, ${\cal
H}_0^{(-)}={\cal H}_-^{(1)}\otimes\cdots\otimes{\cal H}_-^{(d)}$, in which the
spinor representation is given by
$\tilde\gamma_m^+=-\tilde\gamma_m^-\equiv\tilde\sigma_m$ and $p_m=0$ for all
$m=1,\dots,d$. Again we find from \eqn{Hintdef} and \eqn{intvanish} the
vanishing \eqn{Avanish} of internal gauge degrees of freedom. Let
$\Pi_0^{(-)}:{\cal H}\to{\cal H}_0^{(-)}$ be the corresponding orthogonal
projection, and define $\alg_0^{(-)}=\Pi_0^{(-)}\,({\rm
End}_{\overline{\Dirac}}\,\alg)\,\Pi_0^{(-)}$. Then proceeding just as above,
we find that the corresponding low-energy projection of the spectral triple
$(\alg,{\cal H},\Dirac)$ is given by
\be
\left(\alg_0^{(-)}~,~{\cal
H}_0^{(-)}~,~\Dirac~\Pi_0^{(-)}\right)~\cong~\left(C^\infty(T_*^d,\complex)
~,~L^2(T_*^d,\tilde S^-)~,~i\,\tilde{\partial\slash}\right)
\label{specDiso}\ee
where
\be
i\,\tilde{\partial\slash}=-i\,\tilde\sigma_m\otimes
\mbox{$\frac\partial{\partial x_m^*}$}
\label{Drestrict}\ee
and $x_m^*=\frac1{\sqrt2}\,\tilde G_{mn}(x_+^n-x_-^n)\in T_*^d$ with $\tilde
G_{mn}=(G+B)_{ml}\,G^{lp}\,(G-B)_{pn}$ the metric on the dual torus $T_*^d$.
Here $\tilde S^-\to T_*^d$ is the spin bundle induced by the projection onto
${\cal H}_0^{(-)}$. Again we recover an ordinary classical spacetime geometry
with all internal gauge degrees of freedom completely suppressed.

The key point is that the quantum mapping which identifies the distinct
classical spacetimes \eqn{spectripproj} and \eqn{specDiso} is achieved as an
inner automorphism of the {\it entire} noncommutative geometry $(\alg,{\cal
H},D)$. Only after the isomorphism is carried out in the full spectral triple
are the projected low-energy subspaces identified. In the present case, the
transformation $T:{\cal H}\to{\cal H}$ is the $T$-duality transformation of the
heterotic string spacetime which can be read off from the analysis of
\cite{lizzi1} and is given via multiplication of the operators of $\alg$ and
elements of $\cal H$ by the background matrices $d^\pm$. It can be readily
shown to be given by
\bea
T\,\gamma_m^\pm\,T^{-1}&=&G^{ln}\left(G\mp B\pm\mbox{$\frac14$}\,A^L\otimes
A_L\right)_{ml}\,\gamma_n^\pm\label{Tgammam}\\T\,\gamma_M\,T^{-1}&\equiv&
\gamma_M~~{\rm mod}~SL(8k,\zed)\label{TgammaM}\\T\,\alpha_k^{(+)m}\,
T^{-1}&=&\left(G+\mbox{$\frac14$}\,A^L\otimes A_L\right)_{nl}
\left[(G-B)^{-1}\right]^{mn}\,\alpha_k^{(+)l}\label{Talpha+}
\\T\,\alpha_k^{(-)m}\,T^{-1}&=&-\left\{\left(G+\mbox{$\frac14$}
\,A^L\otimes A_L\right)_{nl}\left[(G+B)^{-1}\right]^{mn}-
\mbox{$\frac12$}\left(A^L\otimes A_L\right)^m_l\right\}\,
\alpha_k^{(-)l}\nn\\& &\label{Talpha-}\\T\,\alpha_k^{(+)M}\,
T^{-1}&\equiv&\alpha_k^{(+)M}~~{\rm mod}~SL(8k,\zed)
\label{TalphaM}\eea
The transformations \eqn{TgammaM} and \eqn{TalphaM} are defined up to
multiplication by an element of the mapping class group $SL(8k,\zed)$ of the
self-dual lattice $\Gamma_{\cal G}$. The isomorphism defined by
\eqn{Tgammam}--\eqn{TalphaM} can be interpreted in the usual way
\cite{giveon2,giveon1} in terms of the background fields of the string
sigma-model and are equivalent to inversion $d^\pm\to(d^\pm)^{-1}$ of the
background matrices. Given the extra transformations induced by the internal
subspace ${\cal H}_{\rm int}$ (i.e. by the conditions \eqn{Hintdef}), the full
duality group generated by all mappings amongst the subspaces of the kernels of
the Dirac operators as described in \cite{lizzi1} is isomorphic to
$O(d+8k,d,\zed)$. This is the isometry group of the projected Narain lattice
$\Pi^-\Lambda_{\rm N}$ which is the subgroup of the isometry group
$O(d+8k,d+8k,\zed)$ of \eqn{lattice} that preserves the heterotic structure of
the background matrices. Following \cite{lizzi1,lizzi5} one can construct the
explicit operators $T$ which implement the quantum geometry as inner
automorphisms of the vertex operator algebra $\alg$. In particular, as
mentioned before, the automorphism group of the spectral triple in this way
naturally encodes a representation of the affine gauge symmetry associated with
the internal Lie group $\cal G$.

We have therefore seen that although the quantum geometry of the heterotic
spacetime is determined by the duality group $O(d+8k,d,\zed)$ associated with
the toroidal background $T^d\times T^d_*\times T^{8k}$, the extra $8k$
dimensional left-moving gauge degrees of freedom are unobservable in the
low-energy mappings between the kernels \eqn{kerD} and \eqn{kerbarD}, leaving
the same classical toroidal spectral triples as in the case of an ordinary
bosonic string spacetime \cite{lizzi1}. On the other hand, in the full
noncommutative spacetime, whereby the strings oscillate with very high energies
$\alpha_k\gg1$, the internal gauge degrees of freedom are an intrinsic
ingredient of the quantum geometry. Thus at some distance scale, the gauge
degrees of freedom are excited and appear in the spacetime by what can be
thought of as a Kaluza-Klein mechanism. The interpolation between different
spectral triples gives an exact equivalence between various physical theories
built on different compactification lattices. It turns out that the internal
modes come into play in the tachyon sector which is the smallest perturbation
of the classical spacetime whereby the center of mass modes of the string
oscillate with a winding number $w\neq0$, but whose higher vibrational modes
are still turned off. The tachyon subalgebra of $\alg$ is in turn related to a
particular embedding of a noncommutative torus into the lattice vertex operator
algebra which can be thought of as the intermediate structure separating the
classical spacetime from the full-blown noncommutative geometry. It is here
that one sees explicitly the appearance of the gauge fields and how they vanish
at very low energies. We shall now describe how this works.

The operator products among pairs of the fields \eqn{Vnl} and \eqn{VN} can be
worked out straightforwardly and one finds
\bea
V_{q^+q^-}(z_+,z_-)\,V_{r^+r^-}(w_+,w_-)&=&\left(z_+-w_+\right)^{\left
\langle{\cal P}\{q_m^+\}\,,\,{\cal P}\{r_n^+\}\right\rangle}\,
\left(z_--w_-\right)^{\left\langle{\cal P}\{q_m^-\}\,,\,{\cal P}\{r_n^-\}
\right\rangle}\nn\\& &\times\,\NO V_{q^+q^-}(z_+,z_-)\,V_{r^+r^-}(w_+,w_-)
\NO\label{Vcommrel}\\E_K(z_+)\,E_{K'}(w_+)&=&\left(z_+-w_+\right)^{K_M\,
g^{MN}\,K_N'}\,\NO E_K(z_+)\,E_{K'}(w_+)\NO\label{Ecommrel}
\\E_{K^+}(z_+)\,V_{q^+q^-}(w_+,w_-)&=&\left(z_+-w_+\right)^{\left
\langle{\cal P}\{q_m^+\}\,,\,{\cal P}\{K_M^+\}\right\rangle}\nn
\\& &\times\,\NO E_{K^+}(z_+)\,V_{q^+q^-}(w_+,w_-)\NO
\label{EVcommrel}\eea
Interchanging the orders of the operators in these relations leads to cocycle
relations which can be mapped onto the defining relations of a noncommutative
torus \cite{ncg}. The construction is described explicitly in \cite{lizzi3}. In
the present case, we find for the tachyon subalgebra
\be
{\cal
A}_0\equiv\Pi_0\,\alg\,\Pi_0\cong\Pi^-\left((\torus_\theta^{d+8k})_+
\times(\torus_\theta^{d+8k})_-\right)
\label{tachyonalg}\ee
where $\Pi_0$ is the orthogonal projection onto the $L^2$-subspaces of zero
modes in \eqn{HX}, and $(\torus_\theta^{d+8k})_\pm$ are isomorphic copies of
the noncommutative torus $\torus_\theta^{d+8k}\equiv(T^d\times T^{8k})_\theta$
which are together associated with the full Narain lattice \eqn{lattice}. The
deformation parameter is the antisymmetric matrix $\theta^{\mu\nu}$ which is
determined by the metrics of the bilinear forms appearing in the induced phases
in \eqn{Vcommrel}--\eqn{EVcommrel}. Explicitly, we find that
\be
\theta^{\mu\nu}=\left\{\new{\begin{array}{rrl}
G^{mn}\,{\rm sgn}(n-m)~~~~&,&~~~~\mu,\nu=m,n,~~m\neq
n\nn\\\left(G+\mbox{$\frac14$}\,A^l\otimes A_l\right)^{MN}\,{\rm
sgn}(N-M)~~~~&,&~~~~\mu,\nu=M,N,~~M\neq N\nn\\-\mbox{$\frac12$}\,A^{mN}\,{\rm
sgn}(N-m)~~~~&,&~~~~\mu,\nu=m,N\nn\\0~~~~&,&~~~~\mu=\nu\nn\end{array}}\right.
\label{thetadef}\ee
The deformed product of two projected vertex operators ${\cal V}_0=\Pi_0\,{\cal
V}\,\Pi_0$ and ${\cal W}_0=\Pi_0\,{\cal W}\,\Pi_0$ is given by
\be
{\cal V}_0\star_\theta{\cal W}_0=\Pi_0\,({\cal V}\,{\cal W})\,\Pi_0
\label{defprod}\ee
so that the vertex operator algebra gives a natural way to deform the algebra
of functions on the torus $T^d\times T^{8k}$.

One sees therefore that at the level of the noncommutative torus
$\torus_\theta^{d+8k}$, the internal lattice and gauge degrees of freedom give
a non-trivial mixing between the target space and the internal space of the
strings. From \eqn{g} it follows that at distance scales of the order of
$G_{mn}\gg A^L_mA_{Ln},G_{MN}$, we recover the usual quantum deformation of the
spacetime $T^d$ determined by the metric $G^{mn}$ with all gauge degrees of
freedom suppressed. At very large distances $G_{mn}\to\infty$, we recover the
ordinary classical spacetimes described above. On the other hand, at very short
distance scales where $G_{mn}\ll A^L_mA_{Ln},G_{MN}$, we obtain a deformation
of $T^d\times T^{8k}$ determined entirely by the isospin degrees of freedom.
This shows explicitly that at very short distances the usual target space
Riemannian geometry is deteriorated and distances in the spacetime are measured
in terms of the internal noncommutative geometry. Thus the gauge field degrees
of freedom control features of general relativity at the Planck scale. We will
now use this relationship with the noncommutative torus to analyse some more
quantitative aspects of the gauge induced Planck scale geometry. For ease of
notation we shall henceforth assume that $d$ is even.

The most natural way to study the effects of the isospin degrees of freedom is
to consider a gauge bundle over the tachyon sector of the noncommutative
spacetime. We shall therefore introduce a finitely-generated projective
$\alg_0$-module ${\cal E}_\theta\to\torus_\theta^{d+8k}$ over the
noncommutative torus associated with the left-moving sector of the theory. It
can be characterized as follows. Consider the Lie algebra ${\cal L}_\theta$ of
linear derivations of $\alg_0$ and the corresponding Grassmann algebra
${\mit\Lambda}({\cal L}_\theta^*)$ over its dual vector space. Using the
lattice $\Pi^-\Lambda_{\rm N}\subset{\cal L}_\theta^*$ we may define the
integral part ${\mit\Lambda}(\Pi^-\Lambda_{\rm N})$ of this Grassmann algebra
which inherits a natural $\zed_2$-grading ${\mit\Lambda}(\Pi^-\Lambda_{\rm
N})={\mit\Lambda}^+(\Pi^-\Lambda_{\rm
N})\oplus{\mit\Lambda}^-(\Pi^-\Lambda_{\rm N})$ from the decomposition of
${\mit\Lambda}({\cal L}_\theta^*)$ into its Grassmann even and odd degree
components ${\mit\Lambda}^\pm({\cal L}_\theta^*)$, respectively. The module
${\cal E}_\theta$ may then be identified with an element of the $K$-theory
group $K_0(\alg_0)\cong{\mit\Lambda}^+(\Pi^-\Lambda_{\rm N})$. More precisely,
${\cal E}_\theta$ is identified with its representative $\mu({\cal
E}_\theta)\in K_0^+(\alg_0)$ which lives in the positive cone of $K_0(\alg_0)$
and is considered as an element of the Grassmann-even integer cyclic cohomology
ring $H^+(\torus_\theta^{d+8k},\zed)$.

The Chern character of the gauge bundle may then be represented as \cite{chern}
\be
{\rm ch}({\cal E}_\theta)=\exp(\theta)\,\neg\,\mu({\cal E}_\theta)\in
H^+(\torus_\theta^{d+8k},\real)
\label{cherndef}\ee
where $\neg$ denotes interior multiplication and here the antisymmetric matrix
$\theta$ is considered as a bivector over the Grassmann algebra. Let
$\lambda^\mu$ be a basis of ${\cal L}_\theta^*$. Then the bundle ${\cal
E}_\theta$ can be characterized by a set of integers ${\cal N}$, ${\cal M}$ and
${\cal K}_{\mu_1\dots\mu_{2k}}$ defined by the expansion of its $K$-theory
representative with respect to this basis,
\be
\mu({\cal E}_\theta)={\cal N}+\sum_{i=1}^{\frac12(d+8k-2)}\frac{{\cal
K}_{\mu_1\dots\mu_{2i}}}{(2i)!}\,\lambda^{\mu_1}\wedge\cdots\wedge
\lambda^{\mu_{2i}}+{\cal M}\,\lambda^1\wedge\cdots\wedge\lambda^{d+8k}
\label{mucalEdef}\ee
The corresponding expansion of \eqn{cherndef} then defines the Chern
characteristic classes and the rank of the $\alg_0$-module ${\cal E}_\theta$ as
\be
{\rm ch}({\cal E}_\theta)\equiv\dim{\cal
E}_\theta+\sum_{i=1}^{\frac12(d+8k-2)}\frac{{\rm ch}^{(i)}({\cal
E}_\theta)_{\mu_1\dots\mu_{2i}}}{(2i)!}\,\lambda^{\mu_1}\wedge\cdots
\wedge\lambda^{\mu_{2i}}+{\rm ch}^{(d+8k)}({\cal E}_\theta)\,\lambda^1
\wedge\cdots\wedge\lambda^{d+8k}
\label{chernexp}\ee
After some algebra, we find from \eqn{thetadef} and \eqn{cherndef} the
components
\bea
\dim{\cal E}_\theta&=&{\cal N}+\sum_{i=1}^{\frac12(d+8k-2)}\frac{{\cal
K}_{\mu_1\dots\mu_{2i}}}{(2i)!}\,\theta^{\mu_1\mu_2}\cdots
\theta^{\mu_{2i-1}\mu_{2i}}+{\cal M}\,\det G^{mn}\,\det G^{MN}\nn
\\& &\label{dimE}\\{\rm ch}^{(i)}({\cal E}_\theta)_{\mu_1\dots\mu_{2i}}
&=&{\cal K}_{\mu_1\dots\mu_{2i}}+{\cal M}\left(*\,\theta^{\wedge i}
\right)_{\mu_1\dots\mu_{2i}}\nn\\& &+\sum_{j=i+1}^{\frac12(d+8k-2)}
\frac{[(2i)!]^2}{(2j)!}\,{\cal K}_{\mu_1\dots\mu_{2i}\left[\nu_1
\dots\nu_{2j-2i}\right.}\,\theta^{\nu_1\nu_2}\cdots\theta^{\left.
\nu_{2j-2i-1}\nu_{2j-2i}\right]}\label{chi}\\{\rm ch}^{(d+8k)}
({\cal E}_\theta)&=&{\cal M}
\label{chernnumber}\eea
where $*\,\theta^{\wedge i}$ is the usual Hodge dual defined by viewing
$\theta$ as a bivector on the Grassmann algebra.. The integers $\cal N$, $\cal
K$ and $\cal M$ represent the Chern classes of the undeformed torus $T^{d+8k}$.
We see therefore that the Chern number $\cal M$ of the space is unchanged by
the deformation, but the gauge group rank ${\cal N}_\theta=\dim{\cal E}_\theta$
and the zero modes of the invariant curvature operators $(F^{(0)\,\wedge
i})_{\mu_1\dots\mu_{2i}}$ are shifted by quantities which depend on both the
short-distance spacetime metrics and the internal gauge degrees of freedom. At
very short-distance scales there is therefore a mixing between the various
integers labelling the winding of gauge field configurations around cycles of
the torus.

To get an idea of what this mixing means more precisely, let us consider a
specific vector bundle over the ordinary torus $T^{d+8k}$ whereby only the rank
2 curvature integers ${\cal K}_{\mu\nu}$ are non-vanishing. Then the number of
colours and the field strength zero modes are modified by the deformation
according to
\bea
{\cal N}_\theta&=&{\cal N}+\sum_{m<n}{\cal K}_{mn}\,G^{mn}+\sum_{M<N}{\cal
K}_{MN}\left(G+\mbox{$\frac14$}\,A^l\otimes
A_l\right)^{MN}-\mbox{$\frac12$}\,{\cal K}_{mN}\,A^{mN}\nn\\& &+\,{\cal
M}\,\det G^{mn}\,\det G^{MN}\label{colourshift}\\F_{\mu\nu}^{(0)}&=&{\cal
K}_{\mu\nu}+\frac{2{\cal
M}}{(d+8k-2)!}\,\epsilon_{\mu\nu\nu_1\dots\nu_{d+8k-2}}\,
\theta^{\nu_1\nu_2}\cdots\theta^{\nu_{d+8k-3}\nu_{d+8k-2}}
\label{fieldshift}\eea
where $\epsilon$ is the antisymmetric tensor. Even at very long wavelengths
$G^{mn}\to0$, the colour rank and field strength modes are altered by the
isospin degrees of freedom. In the extreme classical regime whereby only
external target space field modes are excited, i.e. only ${\cal K}_{mn}$ are
non-vanishing, the number of colours remains ${\cal N}$ but the classical
curvatures $F_{mn}^{(0)}$ of $T^d$ are shifted and moreover extra internal
gauge degrees of freedom $F_{mN}^{(0)}$ and $F_{MN}^{(0)}$ are induced by the
internal gauge fields $A:\Gamma_{\cal G}\to\Lambda^*$. This is the essence of
the Kaluza-Klein mechanism, which mixes non-trivially the effective target
space cycles.

What this all means is that the quantum numbers labelling gauge degrees of
freedom depend crucially on the distance scales on which they are observed. The
colour number ${\cal N}_\theta$ depends explicitly on the volume form and
topology of the target space and also on the internal gauge degrees of freedom.
This feature arises because in the tachyon sector of the noncommutative
spacetime the gauge degrees of freedom need to be regarded as elements of
$K$-theory, which thereby produces a non-trivial mixing between internal and
target space modes leading to a Kaluza-Klein inducing of intrinsic gauge
fields. This is consistent with the fact that the lattice vertex operator
algebra associated with $\Pi^-\Lambda_{\rm N}$ naturally contains a non-trivial
projective representation of the homotopy group
$\pi_0(\Omega\,T^{8k})\cong\Gamma_{\cal G}$, where $\Omega\,T^{8k}$ is the loop
space of maps $S^1\to T^{8k}$ corresponding to the internal space. The
two-cocycles \eqn{2cocycle} of this projective representation lead explicitly
to the shifts such as \eqn{colourshift} in the charges of objects which couple
to gauge field excitations. The relevance of loop space geometry for such
noncommutative spaces is discussed in \cite{lizzi2}.\footnote{Note that the
{\it smeared} vertex operator algebra can be regarded as a quantized algebra of
functions on $\Omega(T^d\times T_*^d\times T^{8k})$.} Here we find that it
completely determines the nature of the gauge sector of physical theories which
are constructed in the heterotic spacetime.


\begin{thebibliography}{99}

\baselineskip=12pt

\bibitem{giveon2} A. Giveon, M. Porrati and E. Rabinovici, Phys. Rep. 244
(1994) 77.

\bibitem{ncg} A. Connes, {\it Noncommutative Geometry} (Academic Press, 1994);
G. Landi, {\it An Introduction to Noncommutative Spaces and their Geometries}
(Springer, 1997).

\bibitem{fg} J. Fr\"ohlich and K. Gaw\c edzki, CRM Proc. Lecture Notes 7 (1994)
57; A.H. Chamseddine and J. Fr\"ohlich, in: {\sl Yang Festschrift}, eds. C.S.
Liu and S.-F. Yau (International Press, Boston, 1995) 10.

\bibitem{chams} A.H. Chamseddine, Phys. Lett. B400 (1997) 87; Phys. Rev. D56
(1997) 3555.

\bibitem{lizzi1} F. Lizzi and R.J. Szabo, Phys. Rev. Lett. 79 (1997) 3581;
Comm. Math. Phys. 197 (1998) 667.

\bibitem{susyncg} J. Fr\"ohlich, O. Grandjean and A. Recknagel, {\it
Supersymmetric Quantum Theory, Noncommutative Geometry and Gravitation},
hep-th/9706132, to appear in {\sl Quantum Symmetries}, Les Houches session 64,
eds. A. Connes and K. Gaw\c edzki.

\bibitem{lizzi5} F. Lizzi and R.J. Szabo, {\it Noncommutative Geometry and
Spacetime Gauge Symmetries of String Theory}, hep-th/9712206, to appear in
Chaos, Solitons and Fractals.

\bibitem{lizzi3} G. Landi, F. Lizzi and R.J. Szabo, {\it String Geometry and
the Noncommutative Torus}, hep-th/9806099.

\bibitem{gross2} D.J. Gross, J.A. Harvey, E. Martinec and R. Rohm, Nucl. Phys.
B256 (1985) 253.

\bibitem{narain1} K.S. Narain, Phys. Lett. B169 (1986) 41; K.S. Narain, M.H.
Sarmadi and E. Witten, Nucl. Phys. B279 (1987) 369; P. Ginsparg, Phys. Rev. D35
(1987) 648.

\bibitem{giveon1} A. Giveon, E. Rabinovici and G. Veneziano, Nucl. Phys. B322
(1989) 167.

\bibitem{voa} I.B. Frenkel and V.G. Kac, Invent. Math. 62 (1980) 23; G. Segal,
Comm. Math. Phys. 80 (1982) 301; P. Goddard and D. Olive, Int. J. Mod. Phys. A1
(1986) 303; I.B. Frenkel, J. Lepowsky and A. Meurman, {\it Vertex Operator
Algebras and the Monster}, Pure Appl. Math. 134 (Academic Press, New York,
1988).

\bibitem{lizzi2} F. Lizzi and R.J. Szabo, Phys. Lett. B417 (1998) 303.

\bibitem{chamspec} A.H. Chamseddine, Phys. Lett. B436 (1998) 84.

\bibitem{specaction} A.H. Chamseddine and A. Connes, Phys. Rev. Lett. 77 (1996)
4868; Comm. Math. Phys. 186 (1997) 731.

\bibitem{chern} G.A. Elliot, in: {\sl Operator Algebras and Group
Representations} (Pitman, London, 1984) 157; M.A. Rieffel, Can. J. Math. 40
(1988) 257.

\end{thebibliography}
\end{document}